\documentclass{nature}

\title{Evidence for atomic-scale vibron-mediated electron bunching}

\author{A. Maiti$^{1}$, M. Amato$^{1}$, V. S. Stolyarov$^{2}$, H. Aubin$^{3}$, J. Estève$^{1}$, F. Pistolesi$^{4}$, M. Aprili$^{1}$ \& F. Massee$^{1, *}$}

\usepackage{lineno}
\usepackage[english]{babel}
\usepackage{amssymb,amsmath}
\usepackage{graphicx}
\usepackage{multirow}
\usepackage{bigstrut}

\begin{document}
	
	\maketitle
	
	\begin{affiliations}
		\item Universit\'{e} Paris-Saclay, CNRS, Laboratoire de Physique des Solides, 91405, Orsay, France
		\item Moscow Institute of Physics and Technology, 141700 Dolgoprudny, Russia
		\item Université Paris-Saclay, CNRS, Centre de Nanosciences et de Nanotechnologies, 91120, Palaiseau, France
		\item Université Bordeaux, CNRS, LOMA, UMR 5798, F-33405 Talence, France\\
		* freek.massee@universite-paris-saclay.fr
	\end{affiliations}
	
	\begin{abstract}
		Due to the Coulomb blockade effect, electrons rarely bunch during transport, a phenomenon observed only in a few specially engineered mesoscopic configurations.
		In this work, we introduce an atomically resolved shot-noise study to demonstrate the possibility of electron bunching through vibrational coupling which takes place in an atomically sized nano-electro-mechanical system. Using tunnelling spectroscopy, we observe signatures of vibron-assisted tunnelling on an Fe impurity in Bi$_2$Se$_3$. Notably, simultaneous shot-noise measurements at the centre of the vibrating impurity reveal super-Poissonian noise. In the absence of alternative sources of super-Poissonian noise, this implies vibronic-coupling-induced bunching of electrons during the tunnelling process through the impurity, as theoretically predicted decades ago. As a future outlook, if coherence between electrons can be implemented, vibron-mediated electron bunching at single atomic sites may be exploited as a local injection source of $N$-paired electrons.
	\end{abstract}
	
	\section{Introduction}
	In nanostructures, electronics and mechanics can interfere because of the small energy required to move atoms from equilibrium. As a result, electronic processes occur within a highly fluctuating mechanical environment \cite{Park_NAT, Yu_PRL, Steele_SCIENCE}, which, in the presence of strong coupling to vibronic modes, can deviate significantly from the conventional transport paradigm \cite{Galperin_JPCM, Franke_JPCM}. The common understanding dictates that, as Fermions, electrons inherently obey the Pauli exclusion principle, which prevents them from simultaneously occupying the same quantum state, although two electrons with opposite spins may still share a spin-degenerate orbital. But, Coulomb blockade, the charging energy cost associated with adding an extra electron, restricts occupation to a single electron at a time, even in spin-degenerate states. Therefore, an electron can only enter a localised state when another leaves, creating an `ordered' stream of quasi-particles if the dwell time approaches the average time between tunnelling events -- a phenomenon known as anti-bunching \cite{Korotkov_PRB, Emary_PRB, Birk_PRL}. However, in the presence of strong vibronic coupling and slow vibron relaxation, the opposite phenomenon, electron bunching, has been predicted in both the incoherent \cite{Koch_PRL, Koch_PRB} and coherent \cite{Mitra_PRB,Pistolesi_PRB} transport regimes. The underlying idea is that even long after an electron has tunnelled, the vibron is still active and modulates the tunnelling rate, resulting in positive current correlations. This behaviour not only challenges the conventional notions of atomic-scale electron transport but is also of interest for the on-demand injection of $N$-bunched electrons into quantum matter without complex mesoscopic configurations \cite{Zarchin_PRB, Choi_Nat_Comm, Lau_NL}. Despite its implications, experimental evidence of vibron-mediated electron bunching is sparse \cite{Lau_NL}, and remains elusive at the atomic scale.
	
	\subsection{Principle of the experiment} 
	
\begin{figure}
		\centering
		\includegraphics[width=1\textwidth]{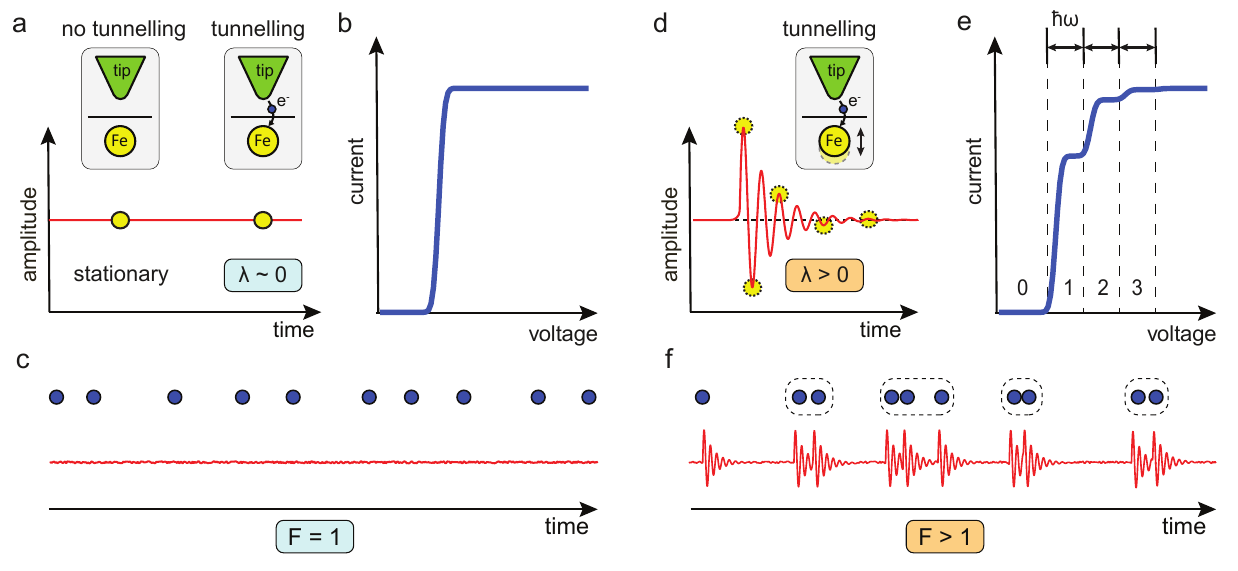}
		
		\caption{
			\textbf{Schematic illustration of the experiment highlighting the key aspects of the study.}
			\textbf{a} A subsurface impurity (Fe) is considered in a solid-state host (Bi$_2$Se$_3$) under the STM tip. In the absence of tunnelling, the impurity remains at rest in a stationary state with only zero-point motion. For weak vibronic coupling ($\lambda \sim 0 $) and/or with fast phonon relaxation, tunnelling is elastic and the vibration effectively remains unexcited.
			\textbf{b} Corresponding current–voltage ($I$–$V$) characteristics exhibit a step-like increase in current without vibrational features.
			\textbf{c} In this case, the tunnelling events are temporally uncorrelated, resulting in Poissonian shot-noise (Fano factor: $F=1$). Nonetheless, anti-bunching behaviour ($F<1$) may emerge when strong Coulomb repulsion prevents electrons from occupying the same energy state simultaneously and transport becomes sequential.
			\textbf{d} For strong electron-vibron coupling ($\lambda > 0$) and slow phonon relaxation, the injection of an electron into the impurity excites the vibrational mode. 
			\textbf{e} Each time a new process involving a higher number of vibrons per tunnelling event is energetically allowed, the current increases in a step-like manner, with steps separated by the vibron energy ($\hbar\omega$). 
			\textbf{f} Schematic illustration of the electron bunching process in the presence of vibronic excitations (dashed boxes). The non-zero motion of the impurity provides a positive correlation between tunnelling electrons, leading to super-Poissonian noise ($F > 1$). 
		}
		\label{fig:1}
	\end{figure}

	To investigate the feedback of vibrational motion on atomic-scale electronic transport, an electronic state strongly coupled to a mechanical resonator is required. An atomic impurity coupled to a vibron (phonon) mode of the host lattice represents the smallest limit of this scenario (Fig.~\ref{fig:1}a). The impurity can be treated as a non-interacting two-level system fluctuating between no-charge and a single electron charge state. In the absence of significant electron-vibron coupling, tunnelling occurs elastically, and the vibrational mode remains unexcited. This leads to a smooth onset of current into the impurity resonance with increasing bias (Fig.~\ref{fig:1}b), and the tunnelling events appear temporally uncorrelated (Fig.~\ref{fig:1}c). However, negative correlations can arise when the Coulomb repulsion is sufficiently strong to prevent two electrons from occupying the same state, resulting in anti-bunching behaviour \cite{Korotkov_PRB, Birk_PRL, Emary_PRB}.
	
	In contrast, when the vibronic coupling is strong (Fig.~\ref{fig:1}d), the tunnelling of electrons into the localised state displaces the equilibrium positions of the harmonic oscillator associated with these two different charge states by an amount larger than the zero-point motion of the oscillator. This leads to the suppression of elastic tunnelling at low bias, when the tunnelling rate is much smaller than both the temperature and the vibron frequency, an effect known as the Franck-Condon (FC) blockade \cite{Franck_TFS, Condon_PR, Koch_PRL, Leturcq_NAT_PHYS}. Inelastic tunnelling, however, is allowed: when the applied voltage is large enough to excite the vibronic mode, the system can overcome the blockade and a current can flow. This results in a step-wise increase in the measured current (Fig.~\ref{fig:1}e) where the width of each step corresponds to the resonating frequency of the mode and the step index corresponds to the number of vibrons added to the mode by the tunnelling process. Prior investigations have demonstrated that conventional scanning tunnelling spectroscopy (STS) effectively probes these steps with high energy and spatial resolution \cite{Qui_PRL, Sun_PRL, Reecht_PRL, Li_NatComm}. However, due to the low bandwidth of a conventional scanning tunnelling microscope (STM), these works provided limited insight into tunnelling dynamics and correlation effects. As a result, observing vibron-mediated electron bunching (Fig.~\ref{fig:1}f) at the atomic scale has been extremely challenging, since such a phenomenon is hidden in the electron tunnelling statistics  \cite{Koch_PRL,Koch_PRB,Novotny_PRL,Haupt_PRB,Kumar_PRL}.
	
	\section{Results}
	Here we provide experimental evidence of electron bunching in an atomic-scale system by employing a home-built STM enabled with dedicated MHz-range cryogenic circuitry \cite{Massee_RSI}. In addition to atomically resolving the tunnelling current, this setup allows for simultaneous measurement of the variance of the current. Specifically, the shot-noise, i.e. the current fluctuations resulting from the discreteness of the electron charge \cite{Blanter_PhysRep}, provides direct insight into the statistics of tunnelling events and atomic-scale electron dynamics. The shot-noise power spectral density is given by: $S_{\text{shot}} = 2e|I|F$, where $e$ is the electron charge, $|I|$ the average tunnelling current, and $F$ the Fano factor; for random or Poissonian tunnelling, $F = 1$ \cite{Blanter_PhysRep}. For a correlated transfer of charges, the noise deviates from Poissonian: reduced for a negative correlation, or anti-bunching ($F<1$), and enhanced for a positive correlation, or bunching ($F>1$). 
	
	\subsection{Fe impurity as a nano-electro-mechanical resonator} 
	
	For this study, Fe impurities embedded within Bi$_2$Se$_3$ have been chosen. Apart from their previously demonstrated role as highly mobile atomic rotors \cite{Desvignes_ACSNano}, one of their advantages is that the impurity states are found in the bulk gap of Bi$_2$Se$_3$, thus remaining relatively isolated \cite{Song_PRB, Stolyarov_APL_2017, Abdalla_PRB}. Furthermore, Bi$_2$Se$_3$ has been shown to possess a very large electro-mechanical coupling constant by angle-resolved photoemission spectroscopy (ARPES) \cite{Kondo_PRL}. Previous research also indicates the existence of different surface vibron (phonon) modes in the host lattice that could be coupled to the electronic states of the impurities \cite{Cheng_PRB_2011}. Moreover, density functional theory (DFT) calculations show that the surface mode has no counterpart in the bulk, where the phonon density of states is gapped \cite{Boulares_SSC}. This may be the reason for the sharp mode observed by Raman spectroscopy, and points to a long vibron relaxation time \cite{Kung_PRB}. The combined effect of strong electron-vibron coupling and slow vibron relaxation makes this system an excellent candidate to investigate the theoretically predicted vibron-mediated electron bunching.
	
	\subsection{Conventional STM/S of the Fe impurity} 
	
\begin{figure}
		\centering
		\includegraphics[width=0.5
		\textwidth]{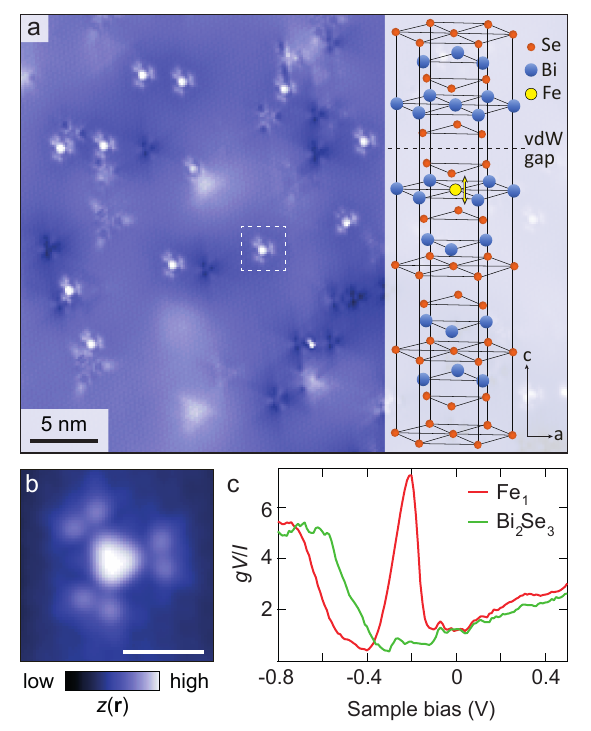}
		
		\caption{
			\textbf{Microscopic and spectroscopic signature of Fe$_1$.} \textbf{a} Large scale topography ($V_{\text{bias}}$ = -0.4 V, $\textit{I}$ = 500 pA) showing a range of different impurities. An isolated Fe$_1$ is marked by the dashed box, its position in the unit cell is shown in the inset: the impurity replaces a Bi atom in the first layer beneath the surface upon cleaving the crystal along the van der Waals (vdW) gap. \textbf{b} A high-resolution constant current image of the Fe$_1$ impurity. A low-pass filter is used to remove 50~Hz noise. The scale bar is 1~nm. (c) Normalised differential conductance taken at a constant current of 500~pA (constant height for $|E|<0.1$~eV) at the centre of the impurity (Fe$_1$) and in an impurity-free area (Bi$_2$Se$_3$). The spectrum at the impurity reveals the presence of a resonance within the bandgap of Bi$_2$Se$_3$}.
		\label{fig:2}
	\end{figure}

	Fig.~\ref{fig:2}a shows a large-scale constant current image of the Se-terminated surface of Bi$_2$Se$_3$ featuring the imprint of different Fe impurities. A comprehensive description of these impurities on different atomic sites, as well as the voltage dependence of their differential conductance, is discussed in Supplementary Information Section 2.1. Primarily, we focus on a single Fe impurity that replaces a Bi atom in the first layer beneath the surface, henceforth called Fe$_1$  (see Fig.~\ref{fig:2}b). The differential conductance ($g = dI/dV$) spectra for a Fe$_1$ impurity and a location on Bi$_2$Se$_3$ in between impurities are presented in Fig.~\ref{fig:2}c. To accurately extract the energy values in the presence of a potential distribution within the double barrier tunnel (DBTJ) junction (tip-impurity-substrate) \cite{Swart_ChemRev}, we measure the differential conductance at a constant current instead of a constant height and normalise it by $I/V$ to remove the set-up effect (see Supplementary Information Section 2.2). The spectra reveal a sharp, occupied resonance peak near the Fermi energy (${E_F}$) at the impurity site, located within the bulk gap of Bi$_2$Se$_3$ \cite{Song_PRB, Stolyarov_APL_2017}. To elucidate the formation of the resonance peak and its orbital contributions, we conducted ground-state DFT calculations (details provided in Supplementary Information Section 1). The analysis reveals that the resonance peak primarily originates from the out-of-plane $d_{z^2}$ orbital, which dominates the total density of states associated with the Fe impurity level. Moreover, the hybridisation of this orbital with the $p_z$ orbitals of the nearest-neighbour surface Se atoms gives rise to the threefold symmetry observed in Fig.~\ref{fig:2}b. This result is particularly intriguing as it suggests the possibility of a vibronic coupling along the direction of tunnelling current, i.e. the out-of-plane direction of the longitudinal $A_{1g}$ phonon modes in Bi$_2$Se$_3$ \cite{Zhang_NL}.
	
	\subsection{Signature of strong vibronic coupling: Franck-Condon principle} 
	
\begin{figure}
		\centering
		\includegraphics[width=1\textwidth]{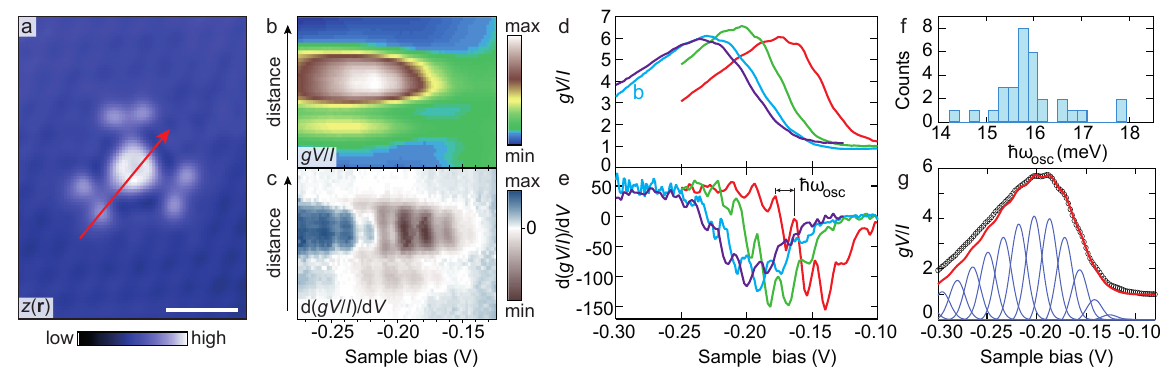}
		
		\caption{
			\textbf{Differential conductance of Fe$_1$.} \textbf{a} Constant current image ($V_{\text{bias}}$ = -0.4 V, $\textit{I}$ = 100 pA) showing an isolated Fe$_1$. A low-pass filter is used to remove 50 Hz noise. The scale bar is 1~nm. \textbf{b} Normalised differential conductance, and \textbf{c} its derivative with respect to voltage taken along the cut indicated by the red arrow in (a), showing clear oscillations. \textbf{d} Characteristic differential conductance spectra and \textbf{e} their derivative taken on different Fe$_1$ impurities. \textbf{f} Histogram of the energy spacing between the maxima of the oscillations, $\hbar \omega_{\text{osc}}$. \textbf{g} Franck-Condon fit of the differential conductance spectra, see Supplementary Information Section 2.4 for fitting details.
		}
		\label{fig:3}
	\end{figure}

	As described earlier (Fig. \ref{fig:1}e), step-like features in the tunnelling current (e.g. peaks in the differential conductance), often referred to as vibronic sidebands, are typically considered a fingerprint of strong vibronic coupling \cite{Qui_PRL, Sun_PRL, Reecht_PRL, Li_NatComm}. Despite its promising characteristics, however, vibron-induced inelastic tunnelling has not previously been observed in Fe-doped Bi$_2$Se$_3$. This may be due to the difference in typical energy scales: whereas the band gap and in-gap resonance energies are on the order of hundreds of meV, vibronic modes are typically in the few to tens of meV range. Therefore, in order to resolve the presence of such fine structures, higher resolution spectra than in Fig.~\ref{fig:2}c are essential. Fig.~\ref{fig:3}a depicts a constant current image of an isolated Fe$_1$ impurity, with Fig.~\ref{fig:3}b showing the voltage-dependent differential conductance taken along the indicated cut. As can be seen, the resonance of Fe$_1$ indeed displays a fine oscillatory structure. While these features are somewhat difficult to resolve in the differential conductance spectra, they are pronounced in their derivative, i.e. in $d(gV/I)/dV$ (Fig.~\ref{fig:3}c). Using a simple fitting procedure with a periodic function (see Supplementary Information Section 2.3), we determine the energy spacing between the periodic oscillations to be $\hbar \omega_{\text{osc}}$ = 16.0 $\pm$ 0.8~meV. Notably, this is indeed in good agreement with the $A^2_{1g}$ surface vibron mode of Bi$_2$Se$_3$ as observed in Raman spectroscopy \cite{Kung_PRB, Zhang_NL} and high-resolution electron energy loss spectroscopy \cite{Jia_PRL}, providing strong support for a vibronic origin. 
	
	To verify that the vibronic sidebands are a common feature of all Fe$_1$ impurities, we conduct a statistical analysis of the characteristic oscillation energies by measuring multiple impurities positioned in diverse local environments. Fig.~\ref{fig:3}d, shows some of the representative data which imply that while the energy position of the resonance peak is significantly affected by the local environment, the energy spacing $\hbar \omega_{\text{osc}}$ remains relatively constant (Fig.~\ref{fig:3}e-f). Furthermore, we can estimate the strength of the dimensionless electron-vibron coupling ($\lambda$) by fitting the differential conductance peaks using the FC factors: \( f_n \propto \frac{{\lambda^{2n}}}{{n!}}\ e^{-\lambda^2}\) with \textit{n} being the difference in vibron quantum numbers \cite{Franck_TFS, Condon_PR}. At equilibrium in the limit of sequential tunnelling, the FC factors give the weight of each multi-vibronic tunnelling process. The differential conductance is then obtained by adding all vibronic sidebands weighted by their respective FC factors, i.e. $g (V) = \sum_n f_n \mathcal{N}(E_n, \sigma)$, where $\mathcal{N}(E_n, \sigma)$ is a Gaussian with a width $\sigma$ that is centred at $E_n = E_0 + n\hbar\omega$. The current $I (V)$ is obtained by integrating $g(V)$, such that we can fit the normalised conductance, $gV/I$ (Fig.~\ref{fig:3}g). We find an average value of $\lambda$ = 2.6 $\pm$ 0.5 which aligns well with a relatively large coupling constant ($\lambda$ $\sim$ 3) previously reported in Bi$_2$Se$_3$ \cite{Kondo_PRL}. 
	
	\subsection{Shot-noise spectroscopy: Exploring electron dynamics} 
	
\begin{figure}
		\centering
		\includegraphics[width=1\textwidth]{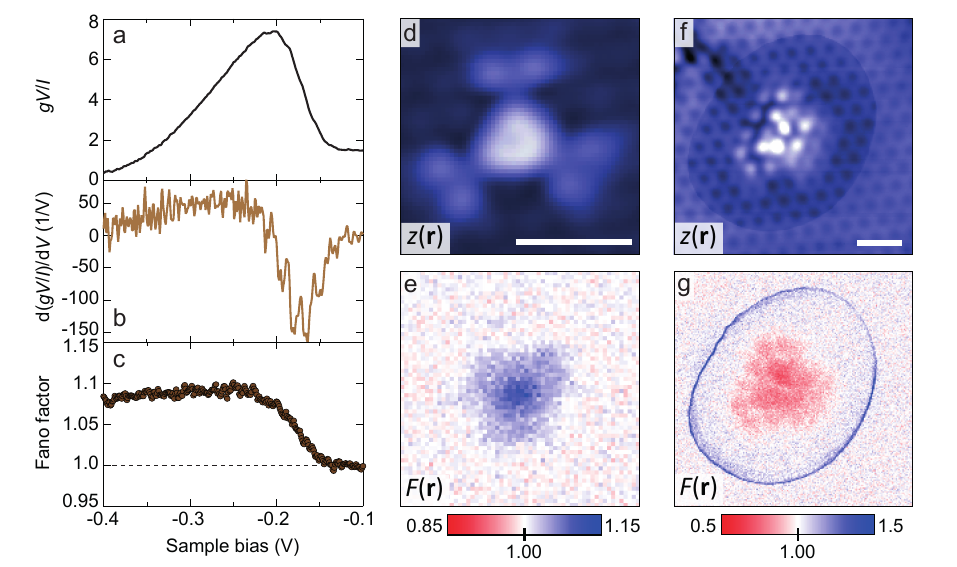}
		
		\caption{\textbf{Visualisation of electron bunching in a vibrating Fe$_1$ and anti-bunching in 2Fe$_1$.} \textbf{a} Normalised differential conductance, \textbf{b} derivative, and \textbf{c} current noise, respectively, taken at $\textit{I}$ = 400 pA at the centre of a Fe$_1$. \textbf{d} Constant current ($\textit{I}$ = 400~pA) image and \textbf{e}, map of the Fano factor extracted from noise measurements on a Fe$_1$. \textbf{f}, \textbf{g} Same as panels (d), (e), but for a 2Fe$_1$ dimer. The scale bar is 1~nm. The noise enhancement observed at the charging ring in (g) follows a non-linear (approximately quadratic) dependence and does not reflect tunnelling statistics or correlation effects, see Fig.~S7, in Supplementary Information Section 2.6. All other noise is linear in current. The colour-scale in panels d, f are the same as in Fig.~\ref{fig:3}a.}
		\label{fig:4}
	\end{figure}

	Having established the generic features of current-induced vibronic coupling in Fe$_1$, we delve into the temporal dynamics of tunnelling events through these impurities. To directly compare the spectroscopic signature and the tunnelling statistics of the vibrating Fe$_1$, we simultaneously measure the differential conductance and current noise ($F=S/2e|I|$) as a function of voltage, while keeping the current fixed. A point spectrum (Fig.~\ref{fig:4}a-c) at the centre of the impurity reveals that the noise follows Poissonian statistics ($F=1$) for small negative voltages, but becomes super-Poissonian ($F>1$) when the voltage is sufficiently high (negative) to tunnel into the resonance. Notice that the Fano factor increases with the number of excited vibrons and saturates around the same energy where the oscillations in the $d(gV/I)/dV$ fade away. The noise then gradually returns to Poissonian near the valence band edge when tunnelling into the substrate becomes predominant (see Supplementary Information Section 2.5). In addition, we observe a (near-) linear current dependence, see Supplementary Information Section 2.6. We next take full advantage of the STM to image the atomically resolved spatial variation of the current noise around the Fe$_1$ impurity. As Figs.~\ref{fig:4}d-e show, the noise is most vividly enhanced at the centre of the Fe impurity below the surface. This strongly suggests that the impurity serves as a focal point for enhanced electron correlations, driven by the dynamics of tunnelling electrons into the localised electronic states of the impurity.
	
	As a control experiment, we extend the noise measurements to other impurities. While most impurities exhibit only weak resonances (see Supplementary Information Section 2.1) that preclude detectable electron dynamics, we identify several nearest-neighbour Fe$_1$ dimers, henceforth termed 2Fe$_1$, all of which show a strong resonance similar to the single Fe$_1$. Intriguingly, although the 2Fe$_1$ is a rarer occurrence and we have fewer statistics than on the Fe$_1$, our studies indicate that the vibronic sidebands appear weaker in 2Fe$_1$ (see Supplementary Information Section 2.7). Strikingly, the noise enhancement observed in Fe$_1$ is absent; instead, the noise is reduced across the dimer site (Fig.~\ref{fig:4}f–g). A previous STM study found a similar suppression of noise for a metallic nanoparticle on an Au substrate \cite{Birk_PRL}. It was argued to be caused by single-electron charging of the nanoparticle, creating negative correlations in the electron transfer. In principle, a similar process could take place in Fe-Bi$_2$Se$_3$: instead of a charging (metallic) nanoparticle, the host material is a semiconductor with reduced mobility at low temperatures, possibly resulting in relatively long relaxation times upon tunnelling into the impurity. The presence of a charging ring at the 2Fe$_1$ (which is absent for the individual Fe$_1$) for energies of the same order of magnitude as the resonance, further supports the interpretation of Coulomb correlation-induced suppression of shot-noise in the 2Fe$_1$. The marked difference between the noise behaviour of the (charging) 2Fe$_1$ and (vibrating) Fe$_1$ provides strong evidence that the noise enhancement in Fe$_1$ is not a spurious effect, but could originate from vibronic coupling as predicted theoretically \cite{Koch_PRL, Koch_PRB, Pistolesi_PRB}.
	
	\section{Discussion}
	
	\subsection{Origin of super Poissonian noise}
	
	Before addressing noise due to vibronic coupling in more detail, we pose the question if the enhanced noise observed on Fe$_1$ can be explained by a different mechanism. The distinct difference in noise for the Fe$_1$ and 2Fe$_1$ on the same sample with the same tip (see also Supplementary Information Section 2.9) rules out numerous alternative sources that could mimic an enhanced noise in Fe$_1$ but are in fact unrelated to shot-noise. These include mechanical instabilities, stochastic charging/discharging, resistance fluctuations, local Joule heating, or simply voltage noise that is converted into current noise by a non-linearity in the density of states. A genuine enhancement of shot-noise in the absence of vibronic motion typically requires multiple tunnelling channels, where the statistics of one conduction channel are affected by a second channel. Examples are resonant tunnelling via interacting levels \cite{Safonov_PRL}, quantum interference between different paths \cite{Cocklin_PRB}, and coupled electron and spin dynamics \cite{Burtzlaff_PRL, Pradhan_PRB}. Our DFT analysis of the single Fe$_1$, consistent with previous work \cite{Abdalla_PRB}, shows that there is only a single peak at negative bias, which has nearly exclusively d$_{z^2}$ character (see Supplementary Information Section 1). Modulated tunnelling is thus unlikely. Moreover, for the 2Fe$_1$ dimer, where the noise is reduced, our DFT calculations show two overlapping peaks at negative bias of mixed in- and out-of-plane character (see Supplementary Information Section 1), effectively ruling out the presence of noise enhancement due to modulated tunnelling. Quantum interference between the tunnelling path of substrate to impurity and substrate to tip is questionable as well, as it would likely also appear in other impurities, whereas we only see enhanced noise on the Fe$_1$ (and reduced noise on the 2Fe$_1$ dimer). Furthermore, the anticipated change in the $F$, in this process \cite{Cocklin_PRB}, is much smaller than the observed enhancement in our experiment. Lastly, as Fe$_1$ has a magnetic moment, its spin dynamics can, in principle, affect the current noise \cite{Burtzlaff_PRL, Pradhan_PRB}. Measurements in an external magnetic field of up to 5T out-of-plane and 1T in-plane, however, did not affect the noise at all (see Supplementary Information Section 2.10).
	
	\subsection{Vibron induced noise}
	
	In the absence of alternative sources of super-Poissonian noise, we revisit vibronic coupling to gain further insight into our noise data. An important observation is the only modest amount of super-Poissonian noise and a linear current dependence of the noise, despite the relatively large electron-phonon coupling constant extracted in this and previous work \cite{Kondo_PRL}. Current-dependent measurements of the differential conductance (see Supplementary Information Section 2.8) also show that while the peak in differential conductance does shift towards more negative energy, $\hbar \omega_{\text{osc}}$ remains constant within error. The absence of significant current dependence, and thus tip-sample distance, suggests that the underlying mechanism for noise enhancement in our case is not the tip-impurity tunnelling rate, but rather 
	the coupling between the displacement of the oscillator and its occupation \cite{Flensberg_NJP}. To model the observed bunching, we therefore use a Holstein coupling between the vibron mode and the impurity, and assume that the tunnelling rate from the impurity to the substrate, $\Gamma$, is much larger than both the tip-impurity tunnelling rate, $\gamma$, and the inverse of the vibron lifetime, $1/\tau$. We introduce the correlation function $g^{(2)}(t)$, which measures the conditional probability of an electron tunnelling at time $t$ given one tunnelled at $t=0$, in direct analogy with the second-order correlation function in optics. Owing to the rate hierarchy, $g^{(2)}(t)$ starts from zero at $t=0$, rises rapidly on a timescale $1/\Gamma$, and reaches a value $g^{(2)}(0^+)>1$. This enhancement reflects the increased tunnelling probability induced by vibron excitation during the first tunnelling event. For instance, the FC factor in the tunnelling rate for the one vibron process is 10 times larger when the vibron mode contains one excitation instead of none (for $\lambda=2.6$). Subsequently, we expect $g^{(2)}(t)$ to relax back to unity over a timescale $\tau$. While more complex dynamics may arise, particularly in the presence of coherence, we neglect them here for simplicity. Within this minimal model, the Fano factor is $F=1+2(I\tau /e)(g^{(2)}(0^+)-1)$.
	
	The linear dependence of the noise on the current (i.e., a constant Fano factor) suggests that $\tau$ is inversely proportional to $I$, implying that the phonon lifetime is limited by inelastic tunnelling $(\tau^{-1}\propto I \propto \gamma)$. In this scenario, however, one would expect a larger Fano factor, given the sizeable value of $\lambda$. An alternative explanation is that the linear behaviour arises from the restricted current range (pA–nA) explored in the experiment. Assuming instead that the phonon lifetime is limited by processes unrelated to tunnelling, and fixing $\tau=1\,\mathrm{ps}$ as reported in THz spectroscopy experiments \cite{Melnikov_PRB}, we obtain $g^{(2)}(0^+)\simeq 13$ in order to reproduce the observed Fano factor of 1.16 (see Supplementary Information Section 3). 
	
	\section{Outlook and Perspectives}
	As a final note, we highlight a few experimental observations that invite future theoretical development. One surprising observation is that the noise on the Fe$_1$ saturates near the energy where sidebands disappear in $d(gV/I)/dV$, which is also where the FC model no longer accurately describes the $gV/I$. Moreover, it then remains enhanced well beyond the resonance and only reduces upon reaching the valence band edge. Theoretical models, considering either vibronic coupling \cite{Koch_PRL, Koch_PRB, Pistolesi_PRB} as well as other mechanisms \cite{Iannaccone_PRL, Safonov_PRL} predict that the noise typically reduces quickly beyond the resonance. Another puzzling observation is the absence of noise enhancement at the Fe$_1$ impurity lobes (see Fig. \ref{fig:4}e), despite the presence of vibronic sidebands (Fig. \ref{fig:3}c). Even with reduced current into the lobes, some enhancement should be detectable (Supplementary Information Section 2.11). This implies that the lateral distance might influence vibronic feedback, likely due to the orbital structure of the impurity and tunnelling paths not considered theoretically. A recent study of FC features  indeed highlighted the importance of including the molecular wave functions for understanding conventional STS \cite{Reecht_PRL}, underlining the need to include them in the theoretical treatment of the noise. More generally, a comprehensive theoretical description of non-equilibrium current noise in the quantum strong-coupling regime ($\lambda \omega_{\rm osc} \gg \Gamma \gg k_B T/\hbar$), relevant for this work, requires further development. On the experimental side, single molecules on top of a metal substrate may offer exciting advantages over our buried impurity. The molecular vibrations will likely have even longer phonon lifetime, and are thus expected to exhibit stronger electron bunching or even avalanche tunnelling. Looking ahead, if coherence among electrons can be established, vibron-induced inelastic tunnelling could not only modify transport statistics, but may also offer a novel route for on-demand injection of $ N$-paired electrons.
	
	\section{Methods}
	\subsection{Samples and measurement details}
	Bi$_2$Se$_3$ single crystals incorporating 0.2\% Fe-impurities were synthesised following a modified vertical Bridgman method. High-purity elements Bi (99.999\%), Se (99.999\%), and Fe (99.998\%) were stoichiometrically weighed according to the nominal composition and sealed under high vacuum (~$10^{-6}$ mbar) in a double-walled quartz ampoule to prevent explosion from Se overpressure. The ampoule was gradually heated to 850 $^\circ$C, homogenised for 48 hours, and then lowered through a steep thermal gradient at a rate of 0.5 mm/h to promote single-crystalline growth. The resulting ingot was then annealed at 450 $^\circ$C for 72 hours to ensure compositional homogeneity and minimise intrinsic crystalline defects.
	
	The single crystal samples were cleaved mechanically in cryogenic vacuum at a temperature of approximately 20~K and subsequently directly placed into the STM head at 4.2~K. All measurements were carried out at $T$ = 1.4~K (unless otherwise specified) using an etched tungsten tip that was characterised on a Pt sample prior to the measurements, achieving stable atomic resolution with energy-independent density of states. Differential conductance data were obtained by using a standard lock-in amplifier with a modulation frequency of 432.1~Hz and a modulation voltage of 2-5~mV. The resulting differential conductance ($g=dI/dV$) was normalised by ($I/V$) to remove the setup effect. STS and simultaneous noise measurements were performed with the home-built setup and MHz circuitry described in Refs. \cite{Massee_RSI, Thupa_PRL}. The noise power in a 20~kHz band around the LC resonant frequency (1.05~MHz) was integrated using a Rohde FSV3000 spectrum analyser after amplification by an NF SA-230F5 voltage amplifier. The noise amplitude spectral density was also measured at the LC resonance using the Nanonis OC4 module, giving identical results.
	
	\subsection {DFT Calculation} 
	DFT calculations were performed using the OpenMX software package (version 3.9) \cite{OzakiPRB2003,OzakiPRB2004}, which employs pseudo-atomic localized basis functions and norm-conserving pseudopotentials. The Perdew–Burke–Ernzerhof (PBE) functional was chosen to approximate the exchange-correlation energy~\cite{PerdewPRL1996}. For Bi\(_2\)Se\(_3\), the basis sets used were Bi8.0-$s^{3}p^{3}d^{2}$ and Se8.0-$s^{3}p^{3}d^{2}$, with a cutoff radius of 8.0 Bohr for both elements. For Fe, the Fe6.0S-$s^{3}p^{2}d^{1}$ atomic basis set was used, with a cutoff radius of 6.0 Bohr. The convergence of the basis sets was verified through test calculations on pristine and doped bulk systems, and the results were in good agreement with previous theoretical works~\cite{Abdalla_PRB, KimPRB2015, PtokJPCM2020}. The experimental lattice parameters of Bi\(_2\)Se\(_3\) (a = 4.14~\AA~and c = 28.64~\AA~for the rhombohedral layer structure~\cite{NakajimaJPCS1963}) were used in all calculations to ensure an accurate computational model. Van der Waals corrections were applied using the method described by S. Grimme \textit{et al.}~\cite{GrimmeJCC2006}. An energy cutoff of 300 Ry was employed for the numerical integration of the charge density and wave functions. The undoped Bi\(_2\)Se\(_3\) slabs were constructed by stacking quintuple layers (QLs) along the c-axis, maintaining the symmetry and cell parameters of the bulk crystal. The slab thickness was fixed to 6 QLs to ensure total energy and surface state convergence. A vacuum region of 20~\AA\ was added to prevent interactions between periodic images in the direction perpendicular to the surface. For the Fe-doped Bi\(_2\)Se\(_3\) slabs, a 6$\times$6$\times$1 supercell (containing 270 atoms) was used to avoid spurious electrostatic interactions between periodic replicas of the dopants. In the $\mathrm{Fe_1}$ configuration, the Fe impurity was inserted by substituting a Bi atom in the topmost QL with a Fe atom, while in the 2$\mathrm{Fe_1}$ configuration, two Fe atoms replaced two Bi first-neighbour atoms in the topmost QL. The inner layers of the slab were fixed in their bulk positions, allowing only the top two layers to relax until the force on each atom was less than 0.0004 Hartree/Bohr. To mitigate numerical difficulties arising from the doped system's metallic nature and to ensure smooth convergence, a non-zero electronic temperature of 500 K was chosen. The energy convergence criterion for electronic self-consistency was set to 10\textsuperscript{-8} eV, ensuring high accuracy in total energy calculations. Brillouin zone (BZ) integration was performed using a k-point mesh of 12$\times$12$\times$1 for the slab calculations. The electronic structure of the Fe-doped Bi\(_2\)Se\(_3\) slabs was obtained by calculating the electronic density of states (DOS), projected onto atomic orbitals (PDOS), with ${E_F}$ set to 0 eV. Spin-orbit coupling was included in all calculations.
	
	\begin{addendum}
		\item[Data Availability] Source data are provided in this paper. Raw data and codes for this work are available on reasonable request.
	\end{addendum}

	\begin{addendum}
		\item[Acknowledgements] We thank W. Belzig, J. C. Cuevas, J. Fransson and S. Pradhan for insightful discussions. A.M. acknowledges funding from HORIZON-MSCA-2023-PF-01 (101152827). F.M. acknowledges funding from the ANR (ANR-21-CE30-0017-01). Fabrication of the samples was supported by the Ministry of Science and Higher Education of the Russian Federation (Agreement No. 075-15-2025-010).
		\item[Author Contributions] A. Maiti and F. Massee carried out the experiments. A. Maiti and F. Massee analysed the experimental data with valuable input from M. Aprili. M. Amato performed the DFT calculations. V. Stolyarov synthesised the samples. J. Estève, F. Pistolesi performed noise calculations. H. Aubin contributed to the interpretation and discussions of the results. A. Maiti, F. Massee, A. Amato and J. Estève prepared the figures and wrote the manuscript with contributions from all authors. F. Massee supervised the project.
		\item[Competing Interests] The authors declare that they have no competing interests.
	\end{addendum}
	
\end{document}